\documentclass[10pt]{article}
\usepackage{epsfig}
\renewcommand{\theequation}{\thesection.\arabic{equation}}
\usepackage{amssymb}

\begin{document}


    \title{\LARGE \bf  On the Origin of Kaluza's Idea of Unification\\}

   \author{
   \large Y. Verbin$^a$\thanks{Electronic address:
verbin@oumail.openu.ac.il} \setcounter{footnote}{3}
  and N.K. Nielsen$^b$\thanks{Electronic address: nkn@fysik.sdu.dk}}
  \date{ }
    \maketitle
    \centerline{$^a$ \em Department of Natural Sciences, The Open
University of Israel}
    \centerline{\em P.O.B. 39328, Tel Aviv 61392, Israel}
     \vskip 0.4cm
    \centerline{$^b$ \em Department of Physics, University of Southern
Denmark, }
    \centerline{\em Campusvej 55, 5230 Odense M, Denmark}

        \vskip 1.1cm

    \begin{abstract}
We argue that the starting point of Kaluza's idea of unifying
electrodynamics and gravity was the analogy between gravitation
and electromagnetism which was pointed out by Einstein and
Thirring. It seems that Kaluza's attention was turned to this
point by the three papers on the Lense-Thirring effect and the
analogy between gravitation and electromagnetism which were
published a short time before Kaluza's paper was submitted. We
provide here an English translation
of the third of these papers.\\
\end{abstract}



\section{Introduction}\label{Introduction}
\setcounter{equation}{0}

In most descriptions of the Kaluza-Klein approach to unification
\cite{O'Raifeartaigh,O'RStr,Wuensch2003,Goenner}, Kaluza's idea
\cite{Kaluza} (English translations exist in
\cite{Lee,AppelquistEtAl}) to add a fifth dimension to spacetime
is considered to be the starting point for the whole structure.

Most authors also mention the previous five-dimensional unifying
theory by Nordstrom \cite{Nordstrom} (English translation in
\cite{AppelquistEtAl}) which was a flat five-dimensional Maxwell
theory with the additional requirement that all dynamical
variables are independent of the fifth coordinate. Nordstrom's
interpretation of the resulting theory was that of a usual
four-dimensional Maxwell system coupled with relativistic scalar
gravity thus obtaining a common source for both kinds of forces.

Since it is generally accepted that Kaluza was unaware of
Nordstrom's theory, Kaluza's idea is regarded as a "quantum leap"
with no direct connection to any previous work except, of course,
Einstein's general relativity. In addition, Weyl's
four-dimensional unifying theory \cite{Weyl} can be considered a
source of the "spirit of unification" of the time.

We wish to suggest a different view, viz. that Kaluza's starting
point for his theory was the analogy between the Maxwell and
Einstein equations for fields of slowly varying sources. This
analogy was pointed out by Thirring (based on earlier work by
Einstein \cite{Einst1913} - English translation in
\cite{EinstColPapV4}) during his work on the Thirring-Lense effect
\cite{Thirring1918a,ThirringLense} (English translation in
\cite{MashhoonEtAl}). It is explicitly mentioned (although in a
footnote) in the first 1918 paper by Thirring \cite{Thirring1918a}
and an expanded treatment appears in a 2-page paper
\cite{Thirring1918b} in the same year.

Indeed, Einstein was the first, in 1913, to note an
electrodynamic-gravitational analogy \cite{Einst1913} but it was
based on a tentative version of the gravitational field equations.
A corrected version which was consistent with the final Einstein
equations is Thirring's 2-page 1918 paper \cite{Thirring1918b}.
This is actually the basis for the gravito-electromagnetism
formalism \cite{Mashhoon2003,TartRugg2003} but it is rarely
acknowledged as such. Usually, the other two Thirring \& Lense
papers (and their translations)  are referred to in this context.

These three papers
\cite{Thirring1918a,ThirringLense,Thirring1918b} were published
slightly before Kaluza prepared a manuscript of his idea and sent
it to Einstein. This must have happened in the first third of 1919
as is evident from Einstein's first very warm reaction in a letter
dated 21 April 1919 \cite{Pais,O'RStr,Wuensch2003,Goenner}.
Moreover, the 2-page 1918 paper by Thirring is cited by Kaluza in
the context of this very same analogy between electrodynamics and
gravity at the beginning of his paper and may be regarded a
preparation and motivation for all the rest.

Therefore, we suggest viewing Kaluza's work as being based on two
main ingredients: the electrodynamic-gravitational analogy and the
addition of a fifth dimension. Kaluza's starting point was the
electrodynamic-gravitational analogy for slowly varying sources.
Next came the observation that this is more than an analogy, in
that general relativity in some sense contains electromagnetism.
Adding a fifth dimension was only the next step which was
unavoidable in order to have enough "room" for both fields within
the same theory. Going to five dimensions was probably an
independent contribution of Kaluza; certainly so was his idea to
go to five-dimensional general relativity.

In order to strengthen our suggestion, in the next section we
provide  an English translation of Thirring's 2-page 1918 paper
\cite{Thirring1918b} so that non-German speaking readers can judge
for themselves. The last section includes a more detailed
discussion about the connection of Thirring's paper to Kaluza's.


\section{English Translation of the Thirring Paper \cite{Thirring1918b}}
  \label{Thirring}
\renewcommand{\theequation}{\arabic{equation}}
\setcounter{equation}{0}

Translated by N. K. Nielsen.

  \vskip 1.1cm

\begin{center}
{\Large \bf On the Formal Analogy between the Electromagnetic
Fundamental Equations and the Einsteinian Equations of Gravity in
the First Approximation}
\end{center}
\begin{center}
by Hans Thirring.
\end{center}
In the following some formal developments will be carried out that
in an earlier article \footnote{H. Thirring, this journal {\bf
19}, 33, 1918.} only found room in a footnote. The matter under
consideration is the analogy between the Maxwell-Lorentz equations
on one hand, and those equations that determine the motion of a
point particle in a weak gravitational field in the first
approximation, on the other. Einstein himself already referred to
this analogy in his speech at the Wiener Naturforschertag 1913
\footnote{A. Einstein, this journal {\bf 14}, 1261, 1913. \\$^*$
Translator's note: literally: Scientist Congress in Vienna 1913.
Actually the reference should be \cite{Einst1913}: A. Einstein,
this journal {\bf 14}, 1249, 1913. The analogy is indeed mentioned
on p. 1261.}$^*$; however, since his field equations have been
subject to quite an important modification, it appears not
improper to develop the formulas in question for the final version
of the theory.

We remark in advance that in the following, we always use a
coordinate system where the velocity of light is 1, and as
coordinates we choose:
$$
x_1=x, x_2=y, x_3=z, x_4=it.
$$

We now consider a rather special case of point motion in a
quasistationary field of gravitation. A mass point moves in this
field so slowly that the squares and products of its velocity
components are negligible compared to 1. The gravitational field
itself is assumed to be weak (such that the deviations of $g_{\mu
\nu }$ from the classical values $-1$ and 0, respectively, can be
considered small quantities of first order) and to be generated by
incoherent (tensionless) moving masses, the velocities of which
are somewhat larger than that of the point mass under
investigation, such that squares and binary products have to be
taken into account. If we denote the velocity of the point mass
$\vec{v}$ and those of the field generating masses
$\vec{v}\hspace{01 mm}'$, we thus have to keep expressions of the
order of magnitude $\mid \vec{v}\mid $, $\mid \vec{v}\hspace{1
mm}'\mid $, $\vec{v}\hspace{1 mm}'^2$ and $\vec{v} \cdot
\vec{v}\hspace{1 mm}'$.

As is well known, the equations of motion are
\begin{equation}
\frac{d^2x_\tau}{ds^2}=\Gamma ^\tau_{\mu \nu }\frac{dx_\mu
}{ds}\frac{dx_\nu }{ds},\hspace{1 mm}\tau=1\cdots 4. \label{one}
\end{equation}
On the right-hand side the velocity components of the point mass
occur; if we neglect, according to our assumptions, their squares
and products the equations become:
\begin{equation}
\frac{d^2x_\tau}{dt^2}=2i(\Gamma ^\tau _{14}\frac{dx_1}{dt}+\Gamma
^\tau _{24}\frac{dx_2}{dt}+\Gamma ^\tau
_{34}\frac{dx_3}{dt})-\Gamma^\tau_{44}. \label{geodaet}
\end{equation}
In the following we consider the three spatial components of the
equations of motion, and thus $\tau =1,2,3.$ For weak fields, the
three-index symbols are:
\begin{equation}
\Gamma^\tau _{\sigma 4}=-\left\{
\begin{array}{cc}
         \sigma  4 \\
          \tau\\
         \end{array}
\right\} =\left[
\begin{array}{cc}
         \sigma  4 \\
          \tau\\
         \end{array}
\right] =\frac 12(\frac{\partial g_{\sigma \tau }}{\partial
x_4}+\frac{\partial g_{\tau 4}}{\partial x_\sigma} -\frac{\partial
g_{\sigma 4}}{\partial x_\tau}).
\end{equation}
The derivatives $\frac{\partial g_{\sigma \tau }}{\partial x_4}$
($\sigma \neq 4, \tau \neq 4$) are, as will turn out immediately,
of order of magnitude $\vec{v}\hspace{1 mm}'^2$; in
(\ref{geodaet}) they are multiplied by $\frac{dx_\sigma}{dt}$
(order of magnitude $\mid \vec{v}\mid $) and are hence negligible
in our approximation. In $\Gamma ^\tau _{44}$ only the derivatives
of $g_{\tau4}$ and $g_{44}$ occur; hence for the following we need
only those coefficients $g_{\mu \nu}$ that contain the index 4 at
least once.  In order to compute them we use the approximate
solution of Einstein \footnote{A. Einstein, Berl. Ber. 1916,
p.688. \\$^*$ Translator's note: more accurately: Sitzungsber.
Konigl. Preuss. Akad. Wiss. (Berlin) {\bf 1916}, 688 (1916).
English translation in \cite{EinstColPapV6}.}$^*$:
\begin{eqnarray}&&
g_{\mu \nu }=-\delta _{\mu \nu }+\gamma _{\mu \nu},\hspace{1 mm}
\delta _{\mu \nu }=\left\{
\begin{array}{cc}
         1, &\mu=\nu,\\
          0,&\mu \neq \nu,\\
         \end{array}
\right. \nonumber\\&& \gamma _{\mu \nu }=\gamma _{\mu \nu }'-\frac
12  \delta _{\mu \nu } \sum _\alpha \gamma _{\alpha \alpha }',
\nonumber\\&& \gamma _{\mu \nu }'=-\frac{\kappa }{2\pi }\int
\frac{T_{\mu \nu}(x',y',z',t'-R)}{R}dV_0, \label{Einstein}
\end{eqnarray}
where $T_{\mu \nu }$ denotes the energy tensor, $x',y',z'$ the
coordinates of the integration space, $R$ is the distance from the
integration element to position of the point mass, and $dV_0$ the
naturally measured volume element. The energy tensor for
incoherent matter is given by
\begin{equation}
T_{\mu \nu }=T^{\mu \nu }=\rho _0\frac{dx_\mu }{ds}\frac{dx_\nu
}{ds}.
\end{equation}
The four relevant coefficients $g_{\mu 4 }$ are accordingly:
\begin{eqnarray}&&
g_{14}=-i\frac{\kappa }{2\pi }\int \frac{\rho
_0v'_x}{R}(\frac{dt'}{ds})^2dV_0, \nonumber\\&&
g_{24}=-i\frac{\kappa }{2\pi }\int \frac{\rho
_0v'_y}{R}(\frac{dt'}{ds})^2dV_0, \nonumber\\&&
g_{34}=-i\frac{\kappa }{2\pi }\int \frac{\rho
_0v'_z}{R}(\frac{dt'}{ds})^2dV_0, \nonumber\\&&
g_{44}=-1+\frac{\kappa }{4\pi }\int \frac{\rho
_0}{R}(\frac{dt'}{ds})^2dV_0. \label{Hilbert}
\end{eqnarray}

The field components $\Gamma ^\tau _{\sigma 4}$ entering
(\ref{geodaet}) are now computed from the $g_{\mu \nu }$ with the
applied approximations as follows:
\begin{eqnarray}&&
\Gamma ^1_{14}=0 \hspace{1 cm} \Gamma ^1_{24}=\frac
12(\frac{\partial g_{14}}{\partial x_2} -\frac{\partial
g_{24}}{\partial x_1}) \hspace{1 cm} \Gamma ^1_{34}=\frac
12(\frac{\partial g_{14}}{\partial x_3} -\frac{\partial
g_{34}}{\partial x_1}) \hspace{1 cm} \Gamma ^1_{44}=-\frac
12\frac{\partial g_{44}}{\partial x_1} +\frac{\partial
g_{14}}{\partial x_4} \nonumber\\&& \Gamma ^2_{14}=\frac
12(\frac{\partial g_{24}}{\partial x_1}-\frac{\partial
g_{14}}{\partial x_2}) \hspace{1 cm} \Gamma ^2_{24}=0 \hspace{1
cm} \Gamma ^2_{34}=\frac 12(\frac{\partial g_{24}}{\partial
x_3}-\frac{\partial g_{34}}{\partial x_2}) \hspace{1 cm} \Gamma
^2_{44}=-\frac 12\frac{\partial g_{44}}{\partial
x_2}+\frac{\partial g_{24}}{\partial x_4} \nonumber\\&& \Gamma
^3_{14}=\frac 12(\frac{\partial g_{34}}{\partial
x_1}-\frac{\partial g_{14}}{\partial x_3}) \hspace{1 cm} \Gamma
^3_{24}=\frac 12(\frac{\partial g_{34}}{\partial
x_2}-\frac{\partial g_{24}}{\partial x_3}) \hspace{1 cm} \Gamma
^3_{34}=0 \hspace{1 cm} \Gamma ^3_{44}=-\frac 12\frac{\partial
g_{44}}{\partial x_3}+\frac{\partial g_{34}}{\partial x_4}
\nonumber\\&& \Gamma ^4_{14}=\frac 12\frac{\partial
g_{44}}{\partial x_1}-\frac{\partial g_{14}}{\partial x_4}
\hspace{1 cm}\Gamma ^4_{24}=\frac 12\frac{\partial
g_{44}}{\partial x_2}-\frac{\partial g_{24}}{\partial x_4}
\hspace{1 cm} \Gamma ^4_{34}=\frac 12\frac{\partial
g_{44}}{\partial x_3}-\frac{\partial g_{34}}{\partial x_4}
\hspace{1 cm} \Gamma ^4_{44}=0. \label{Christoffel}
\end{eqnarray}
The equations (\ref{geodaet}), (\ref{Hilbert}) and
(\ref{Christoffel}) now correspond, disregarding some numerical
factors, completely to the fundamental electrodynamic equations.
In order to make this similarity more obvious, we set
\begin{eqnarray}&&
{\cal A}_x=ig_{14},\hspace{1 cm} {\cal A}_y=ig_{24},\hspace{1
cm}{\cal A}_z=ig_{34},\hspace{1 cm}\Phi=\frac{g_{44}+1}{2}
\nonumber\\&& {\cal H}_x=2i\Gamma ^3_{24}=-2i\Gamma
^2_{34},\hspace{1 cm} {\cal H}_y=2i\Gamma ^1_{34}=-2i\Gamma
^3_{14}, \hspace{1 cm} {\cal H}_z=2i\Gamma ^2_{14}=-2i\Gamma
^1_{24}, \nonumber\\&& {\cal E}_x=\Gamma ^1 _{44}=-\Gamma
^4_{14},\hspace{1 cm} {\cal E}_y=\Gamma ^2 _{44}=-\Gamma
^4_{24},\hspace{1 cm} {\cal E}_z=\Gamma ^3 _{44}=-\Gamma
^4_{34},\hspace{1 cm} \nonumber\\&& k=\frac{\kappa}{8\pi }.
\label{Riemann}
\end{eqnarray}

In terms of these quantities, equations (\ref{geodaet}),
(\ref{Hilbert}) and (\ref{Christoffel}) become:
\begin{eqnarray}&&
\vec{\cal A}=4k\int \frac{\rho _0\vec{v}\hspace{1
mm}'}{R}(\frac{dt\hspace{1 mm}'}{ds})^2dV_0, \nonumber\\&&
\Phi=k\int  \frac{\rho _0}{R}(\frac{dt\hspace{1 mm}'}{ds})^2dV_0,
\hspace{3 cm}(6a) \nonumber
\end{eqnarray}
\begin{eqnarray}
\vec{\cal H}={\rm curl }\vec{\cal A},\hspace{1 cm}\vec{\cal E}
=-{\rm grad}\Phi-\frac{\partial \vec{\cal A}}{\partial t},
\hspace{2 cm}(7a) \nonumber
\end{eqnarray}

\begin{eqnarray}
\stackrel{\ddot \rightarrow}{\cal S}=-\vec{\cal E}-[\vec
v\vec{H}]. \hspace{5 cm}(2a)\nonumber
\end{eqnarray}


Apart from the factor $(\frac{dt'}{ds})^2$ that only deviates from
unity by quantities of order $\vec{v}\hspace{1 mm}'^2$, equations
(6a), (7a) and (2a) only differ from the corresponding
electrodynamic equations in the wrong sign on the right-hand side
of (2a) and in the emergence of a factor 4 in (6a). Thus the
analog of the magnetic force in the theory of gravitation is four
times larger than in electrodynamics.

To the derivation of this formal analogy, a remark of a principal
nature is added. It seems {\it a priori} very unlikely that
mathematical laws that in one area of phenomena are approximated
formulas for certain special cases, provide an exact description
of the phenomena in another area. Thus, the conjecture arises
(apart from the physical necessity, for formal reasons as well)
that the Maxwell-Lorentz equations are also approximate formulas
that, even though they are sufficiently precise for the fields
generated electrotechnically, need a corresponding generalization
for the far stronger fields that occur at the dimensions of atoms
and electrons, to which Hilbert and Mie  (who have a far more
general starting point) have already provided suggestions.

\vspace{.4 cm}

Institute of Theoretical Physics of the University of Vienna,
March 1918.

\begin{flushright}{\small {Received March 26, 1918}}
\end{flushright}

\section{Discussion}
\renewcommand{\theequation}{\thesection.\arabic{equation}}
\setcounter{equation}{0}

First we correct an error which has no effect on the final
conclusion: In eq. (\ref{Christoffel}) for the Christoffel symbols
(see e.g. Wald \cite{Wald} p. 36), there is a mistake in the last
line where the terms $-\frac{\partial g_{i4}}{\partial
x_4},\hspace{1 mm}i=1,2,3,$ should be left out. This also means
that the third line of (\ref{Riemann}) should be corrected since
the equality between the Christoffel symbols does not hold.

\vspace{.4 cm}

It is obvious that the Thirring paper had a strong influence on
Kaluza. Both papers are limited to the weak field limit, they both
use similar methods to isolate and identify the electromagnetic
components and they use an identical matter source - pressureless
dust.

Eqs. (\ref{Christoffel}) and (\ref{Riemann}) could be Kaluza's
starting point. Here Thirring records the gravitational field in
the weak-field limit according to the final version of Einstein's
general theory of relativity of a dust cloud with slow but
otherwise arbitrary motion. It is very likely that Kaluza had
these equations (\ref{Christoffel}) and (\ref{Riemann}) in mind,
when making the conjecture in his paper (second page in both
translations or in the original version) that the electromagnetic
field strength should be "equal to somehow amputated three-index
symbols".

There are of course differences due to the giant (if not
"quantum") leap Kaluza made. A small one is the condition $|g|=
-1$ used by Kaluza but not by Thirring. A more significant
difference is that Kaluza's $x^5$ (actually, he used $x^0$ for the
fourth spacelike coordinate and $x^4$ was $i\times$time) takes the
role of time in the Thirring-Lense papers, so the interpretation
is, accordingly, different. The four-dimensional analogue of
Kaluza's work is the analysis of the gravitational field of a
source which has a weak dependence on one spatial coordinate. The
major difference is, of course, the extra dimension which Kaluza
added. Possibly Kaluza, aware of the "spirit of unification" of
the time, realized that the $D=4$ Maxwell-Einstein analogy
appearing in the Thirring-Lense papers can be turned into a
unifying scheme if a fifth dimension is added.

\vspace{.6 cm}
 {\Large{\bf Acknowledgements}}

\vspace{.4 cm}

We are grateful to F. W. Hehl and D. Wuensch for helpful
 discussions, suggestions and advice.


  \end{document}